\documentclass[aps,prd,twocolumn,showpacs,amsmath,amssymb]{revtex4}

\usepackage{amssymb}
\usepackage{mathrsfs}
\usepackage{txfonts}

\usepackage{graphicx}
\usepackage{dcolumn}
\usepackage{bm}

\begin{document}

\title{Consistency of a kind of general noncanonical warm inflation}

\author{Xiao-Min Zhang\textsuperscript{1}}
\thanks{Corresponding author}
\email{zhangxm@mail.bnu.edu.cn}

\author{Ang Fu\textsuperscript{2}}
\thanks{Corresponding author}
\email{fuang@mail.ustc.edu.cn}

\author{Kai Li\textsuperscript{1}}
\author{Qian Liu\textsuperscript{1}}
\author{Peng-Cheng Chu\textsuperscript{1}}
\author{Hong-Yang Ma\textsuperscript{1}}
\author{Jian-Yang Zhu \textsuperscript{3}}

\affiliation{\textsuperscript{1}School of Science, Qingdao University of Technology, Qingdao 266033, China\\ \textsuperscript{2}School of Mathematical Sciences, University of Science and Technology of China, Hefei 230026, China \\ \textsuperscript{3}Department of Physics, Beijing Normal University, Beijing 100875, China}

\date{\today}

\begin{abstract}
The framework of a kind of noncanonical warm inflation is introduced, and the dynamical equations of this scenario are presented. We propose the slow roll approximations and give some redefining slow roll parameters in this scenario which remain dimensionless. Performing systemic stability analysis, we calculate the slow roll conditions to guarantee that slow roll approximations hold. The slow roll conditions suggest slow roll inflation in general noncanonical warm inflationary scenario can still exist, and in addition, the slow roll approximations are more easily to be satisfied. Then, a concrete Dirac-Born-Infeld warm inflationary model is studied.

\end{abstract}
\maketitle

\section{\label{sec:level1}Introduction}
Inflation, as a necessary supplement to the standard model of cosmology, can solve the problems of horizon, flatness and monopole. Till mow, there are two candidates for inflation: standard inflation, or sometimes called cold inflation, and warm inflation. Cold inflation was first proposed in 1981 \cite{Guth1981,Linde1982,Albrecht1982}, which suppose the inflaton is isolated during inflation, so the Universe is supercool and need a separate reheating phase to warm the Universe. While in warm inflationary Universe, the inflaton is not isolated and instead, it has interactions with other subdominated fields \cite{BereraFang,Berera2000,Berera2006}. The Universe is hot during the whole inflationary phase, so it has an elegant exit and can go smoothly to the radiation-dominated big-bang phase. Cold inflation can produce seeds to give rise to the large scale structure and the observed little anisotropy of cosmological microwave background (CMB) \cite{WMAP,PLANCK} through vacuum fluctuation \cite{Bassett2006,LiddleLyth,Weinberg,Dodelson}, while the density fluctuations naturally generate from thermal fluctuations in warm inflation. Warm inflation can cure ``$\eta$-problem'' \cite{etaproblem} and the overlarge amplitude of the inflaton suffered in cold inflation \cite{Berera2006,BereraIanRamos}. What's more, warm inflation has more relaxed slow roll conditions compared to cold inflation \cite{Ian2008,Campo2010,ZhangZhu,Zhang2014}. Warm inflation can broaden the scope of inflation, which means some models that were ruled out in cold inflation such as the quartic chaotic potential model, able to again be in a very good agreement with the Planck results in warm inflationary scenario \cite{Sam2014}. The swampland conjecture \cite{swamp1,swamp2,swamp3,swamp4,swamp5} proposed recently which may cause conflicts in slow roll cold inflation, can be consistent with warm inflationary models \cite{swampwarm1,swampwarm2}. Some related works reach the conclusion that very small amounts of dissipation can result in warm inflation, so the warm inflation can describe the very early Universe more realizably.

Warm inflation usually takes advantages of canonical field with the Lagrangian density $\mathcal{L}=X-V$, where $X=\frac12 g^{\mu\nu}\partial_{\mu}\phi\partial_{\nu}\phi$ is the kinetic term and $V$ is the potential term, as the inflaton that drives the inflation. Most works about warm inflation concentrate on the model building, the dissipative microphysical mechanism and the cosmological perturbation research of canonical warm inflation \cite{BereraIanRamos,Lisa2004,Mar2007,Mar2013,naturalwarm1,naturalwarm2,Chris2009,Taylor2000,ZhangZhu}. In recent years, kinds of special noncanonical warm inflationary models have emerged, such as warm Dirac-Born-Infeld (DBI) inflation \cite{Cai2011}, warm tachyon inflation \cite{Herrera2006,TachyonZhang,Xiao2011} and warm k-inflation \cite{Peng2018}. As the case in cold inflation, noncanonical inflation also deserve research in warm inflationary picture. We concentrate on the extension of warm inflationary theory for years. At first, we construct the noncanonical warm inflation with a relatively simple Lagrangian density having separable form of kinetic and potential terms \cite{Zhang2014,Zhang2015}. Then we extend warm inflation to the more general noncanonical field with a Lagrangian density having coupling form of kinetic and potential terms \cite{Zhang2018,Zhang2019}. In noncanonical warm inflation, small sound speed and strong dissipation effect can both influence the perturbation observable quantities such as power spectrum, spectral index and tensor-to-scalar ratio etc. \cite{Zhang2014,Zhang2018}. Thanks to the thermal effect and the noncanonical effect, the energy scale when horizon crossing is depressed, which is good news to the assumption that the universe inflation can be described well by effective field theory \cite{Zhang2014,Zhang2018}. And due to the two effects, the magnitude of primordial gravitational wave is weaker than canonical warm inflation and cold inflation. The non-Gaussianity generated in noncanonical warm inflation is studied and found that large noncanonical effect (in other words, a small sound speed) and strong thermal dissipation effect can enhance the magnitude of non-Gaussianity \cite{Zhang2019}. The noncanonical warm inflationary models can safely lie on the allowed region of observations over a wide parameter range. In original warm inflation, it is found that the slow roll approximations was more easily to be satisfied \cite{Ian2008,Campo2010}. The construction of general noncanonical warm inflation broad the scope of inflation theory. However, the consistency issue in general warm inflation is still blank and deserve complete analysis, which is the main task of this paper.

The paper is organized as follows: In Sec. \ref{sec:level2}, we introduce general noncanonical warm inflationary scenario, and give frame equations containing slow roll approximations and redefining slow roll parameters. We make a fully linear stability analysis to obtain slow roll conditions that guarantee the slow roll approximations are valid in Sec. \ref{sec:level3}. A concrete examples in the general warm inflationary scenario is studied in Sec. \ref{sec:level4}. Finally, we summarize our results in Sec. \ref{sec:level5}.

\section{\label{sec:level2}The framework of general noncanonical warm inflation}

In warm inflationary picture, the Universe is a multi-component system, thus the total matter action can be given as:
\begin{equation}\label{action}
S=\int d^4x \sqrt{-g}  \left[\mathcal{L}(X',\varphi)+\mathcal{L}_R+\mathcal{L}_{int}\right],
\end{equation}
where $X'=\frac12g^{\mu\nu}\partial_{\mu}\varphi\partial_{\nu}\varphi$. The Lagrangian density of the noncanonical field is $\mathcal{L}_{noncan}= \mathcal{L}(X',\varphi)$ in above equation, which can be an arbitrary function of the inflaton field $\varphi$ and the kinetic term $X'$. $\mathcal{L}_R$ is the Lagrangian density of the radiation fields and $\mathcal{L}_{int}$ denotes the interaction term between inflaton and other subdominated fields in warm inflationary epoch. There are two conditions that a workable noncanonical Lagrangian density should meet: $\mathcal{L}_{X'}\geq0$ and $\mathcal{L}_{X'X'}\geq0$ \cite{Franche2010,Bean2008}, which means we have $\mathcal{L}_{X'}>1$ in a normalized field representation \cite{Zhang2014}. By convention, the subscripts $\varphi$ and $X'$ denote a derivative.

The motion equation of the inflaton can be obtained by varying above action:
\begin{equation}\label{vary}
  \frac{\partial(\mathcal{L}(X',\varphi)+\mathcal{L}_{int})}{\partial\varphi}-\left(\frac{1}{\sqrt{-g}}\right)
  \partial_{\mu}\left(\sqrt{-g}\frac{\partial\mathcal{L}(X',\varphi)}{\partial(\partial_{\mu}\varphi)}\right)=0.
\end{equation}
In the flat Friedmann-Robertson-Walker (FRW) Universe, the field is homogeneous, i.e. $\varphi=\varphi(t)$, so the motion equation of the inflaton field is reduced to:
\begin{eqnarray}\label{EOM1}
  \left[\left(\frac{\partial\mathcal{L}(X',\varphi)}{\partial X'}+2X'\left(\frac{\partial^2\mathcal{L}(X',\varphi)}{\partial X'^2}\right)\right)\right]\ddot\varphi\nonumber\\+\left[3H\left(\frac{\partial\mathcal{L}(X',\varphi)}{\partial X'}\right)+\dot\varphi\left(\frac{\partial^2\mathcal{L}(X',\varphi)}{\partial X'\partial\varphi}\right)\right]\dot\varphi\nonumber\\-\frac{\partial(\mathcal{L}(X',\varphi)+
  \mathcal{L}_{int})}{\partial\varphi}=0,~~~~~~~~~~~~~~~~~~
\end{eqnarray}
where $H$ is the Hubble parameter.

The energy density of the inflaton is: $\rho(\varphi,X')=2X'\frac{\partial\mathcal{L}}{\partial X'}-\mathcal{L}$, and the pressure is given by: $p(\varphi,X')=\mathcal{L}$. The sound speed is $c_s^2=\frac{\partial p/\partial X'}{\partial\rho/\partial X'}=\left(1+2X'\frac{\mathcal{L}_{X'X'}}{\mathcal{L}_{X'}}\right)^{-1}$, describing the traveling speed of scalar perturbations. The Friedmann equation is
\begin{equation}\label{Hubble}
  H^2=\frac{\rho}{3M_p^2}.
\end{equation}
Considering warm inflationary assumptions, the motion equation of general noncanonical warm inflation can be given by \cite{Zhang2018}:
\begin{equation}\label{EOM2}
  \mathcal{L}_{X'}c_{s}^{-2}\ddot{\varphi}+(3H\mathcal{L}_{X'}+\tilde{\Gamma})\dot{\varphi}+
  \mathcal{L}_{X'\varphi}\dot\varphi^2+V_{eff,\varphi}(\varphi,T)=0,
\end{equation}
where $\tilde{\Gamma}$ is the dissipative coefficient in warm inflation and $V_{eff}$ is the thermal effective potential. In what follows, we will write $V_{eff}$ as $V$ for simplicity. Because of the existing kinetic potential coupling term of the general noncanonical Lagrangian density, the above motion equation has an annoying coupling term $\mathcal{L}_{X'\varphi}\dot\varphi^2$. This brings difficulties in solving the kind of more general warm inflation, especially giving slow roll approximations and calculating perturbations. Fortunately, this problem usually can be solved by making a field redefinition $\phi=f(\varphi)$ to eliminate the coupling term \cite{Zhang2018}. The solutions can be worked out in many cases using analytical or sometimes numerical method. Then the Lagrangian density becomes $\mathcal{L}(X,\phi)$, where $X=\frac12\dot\phi^2$ in the new field representation. Under the field redefinition, an appropriate field representation $\phi=f(\varphi)$ is selected out to make the coupling term disappeared through
\begin{equation}\label{coupling}
  \mathcal{L}_{X\phi}=\frac{1}{f^4_{\varphi}}[f_{\varphi}\mathcal{L}_{X'\varphi}-
  2f_{\varphi\varphi}\mathcal{L}_X'-2f_{\varphi\varphi}\mathcal{L}_{X'X'}X']=0,
\end{equation}
where $f_{\varphi}$ and $f_{\varphi\varphi}$ denote the first- and second-order derivative of the function $f(\varphi)$ respectively. In the redefined uncoupling $\phi$ representation, the slow roll approximations in general noncanonical warm inflation can be given out clearly, and we can get its slow roll conditions by making stability analysis.

In uncoupling $\phi$ representation, the motion equation of the noncanonical inflaton can be expressed as:
\begin{equation}\label{EOM3}
   \mathcal{L}_{X}c_{s}^{-2}\ddot{\phi}+(3H\mathcal{L}_{X}+\Gamma)\dot{\phi}+V_{eff,\phi}(\phi,T)=0.
\end{equation}
$\Gamma$ in above equation is the new dissipative coefficient in the $\phi$ representation and it has the relation $\tilde{\Gamma}=f_{\varphi}^2\Gamma$ with $\tilde{\Gamma}$ \cite{Zhang2018}. With the benefit of field redefinition, the researches on cosmological perturbations in general noncanonical warm inflation can be easy to handle to some extent \cite{Zhang2018,Zhang2019}. Also, we will perform systemic stability analysis using the easy-to-use $\phi$ representation in the following.

The entropy production equation is given by
\begin{equation}\label{entropy}
  T\dot{s}+3HTs=\Gamma\dot{\phi}^{2},
\end{equation}
where $s$ is the entropy density in warm inflation.

The noncanonical inflaton and the dissipative coefficient may not have the ``right'' conventional mass dimension in general noncanonical warm inflation, different to canonical warm inflation. Given this situation, we redefine the dimensionless dissipation strength parameter $r$ in new general noncanonical warm inflation as \cite{Zhang2018}:
\begin{equation}\label{r}
  r=\frac{\Gamma}{3H\mathcal{L}_X}.
\end{equation}
$r\gg1$ suggests that warm inflation is in strong regime, while $r\ll1$ is corresponding to weak regime in warm inflation.

We propose the probable slow roll approximations of general noncanonical warm inflation by dropping the highest derivative terms in the motion equations under the $\phi$ representation. Thus the motion equation of noncanonical inflaton and entropy production equation can be reduced to:
\begin{equation}\label{SREOM}
  3H\mathcal{L}_{X}(1+r)\dot{\phi}+V_{\phi}=0,
\end{equation}
\begin{equation}\label{SRentropy}
  3HTs=\Gamma \dot\phi^2.
\end{equation}
The validity of these slow roll approximations depends on the concrete slow roll conditions in general noncanonical warm inflation, which we will give out in next section. Slow roll conditions of inflation are usually characterized by slow roll parameters. The slow roll parameters in conventional warm inflation are usually defined as:
\begin{equation}
\tilde{\epsilon}=\frac{M_p^2}{2}\left(\frac{V_{\phi}}{V}\right) ^2, \tilde{\eta}=M_p^2\frac {V_{\phi \phi}}{V},
\tilde{\beta}=M_p^2\frac{V_{\phi}\Gamma_{\phi}}{V\Gamma}.
\end{equation}
Typically, above slow roll parameters are no longer dimensionless in general kind noncanonical warm inflation, different from canonical cold and warm inflation. In general noncanonical warm inflation, we define some new slow roll parameters that remain dimensionless for convenience:
\begin{equation}
\epsilon=\frac{M_p^2}{2\mathcal{L}_X}\left(\frac{V_{\phi}}{V}\right) ^2,~~~\eta=\frac{M_p^2}{\mathcal{L}_X}
\frac {V_{\phi \phi}}{V},~~~\beta=\frac{M_p^2}{\mathcal{L}_X}\frac{V_{\phi}\Gamma_{\phi}}{V\Gamma}.
\end{equation}
Now the ``new'' slow roll parameters are all dimensionless, as the ``old'' slow roll parameters in canonical warm inflation. A good feature for the ``new'' slow roll parameters is their invariance under field transformations.

Describing temperature influence in warm inflation, two additional slow roll parameters are given by:
\begin{equation}
b=\frac {TV_{\phi T}}{V_{\phi}},
\end{equation}
and
\begin{equation}
c=\frac{T\Gamma_T}{\Gamma}.
\end{equation}

Then the number of e-folds in general noncanonical warm inflation can be obtained:
\begin{equation}\label{efold}
 N=\int H dt=\int\frac{H}{\dot{\phi}}d\phi\simeq-\frac{1}{M_p^2}\int_{\phi_{\ast}}
^{\phi_{end}}\frac{V\mathcal{L}_X(1+r)}{V_{\phi}}d\phi,
\end{equation}
where $M_p^2=\frac 1{8\pi G}$ and the subscript $\ast$ denoting the time of Hubble horizon crossing.

\section{\label{sec:level3}Stability analysis}

In order to perform a systematic stability analysis conveniently, we define $u=\dot{\phi}$ and then the Eqs. (\ref{EOM3}) and (\ref{entropy}) can be rewritten as:
\begin{equation}\label{EOM4}
     \dot u=-c_s^2\left[3H(1+r)u+\frac{V_{\phi}(\phi,T)}{\mathcal{L}_X}\right],
 \end{equation}
\begin{equation}\label{entropy2}
    \dot{s}=-3Hs+\frac{\Gamma u^2}{T}.
\end{equation}

Inflation is generally associated and dealt with slow roll approximations, and it is the same in general warm inflationary picture. We have proposed its slow roll approximations in Sec. \ref{sec:level2}. The slow roll approximations indicate that the energy density is dominated by potential, the production of radiation is quasi-static, and the evolution of inflaton field is quite slow.

For convenience, $u_0$, $\phi_0$ and $s_0$ is used to express slow roll solutions satisfying these slow roll equations:
\begin{equation}\label{EOM5}
   3H\mathcal{L}_{X}(1+r)u_0+V_{\phi}(\phi,T)=0,
\end{equation}
\begin{equation}\label{entropy3}
    3H_0T_0s_0=\Gamma u_0^2.
\end{equation}

The variables $u$, $\phi$ and $s$ denote the exact solutions satisfying the exact inflationary equations (\ref{Hubble}), (\ref{EOM4}), (\ref{entropy2}). We expand the exact solutions around the slow roll solutions as: $u=u_0+\delta u,$
$\phi=\phi_0+\delta\phi$ and $s=s_0+\delta s$, where the perturbation terms $\delta u,$ $\delta\phi$, and $\delta s$ are much smaller than the background ones $u_0$, $\phi_0$ and $s_0$.

In order to find out conditions to guarantee that slow roll solutions can really act as formal attractor solutions for general warm inflationary dynamical system, we perform stability analysis around the slow roll solutions.
Using the new variable $u$, we can rewrite $X=\frac12u^2$, then $\delta X= u\delta u$, and $\delta\mathcal{L}_{X}=\mathcal{L}_{XX}u\delta u$. With the help of the thermal relation $s=-V_T$, we get $\delta s=-V_{TT}\delta T-V_{\phi T}\delta \phi$. The relation
\begin{equation}\label{deltaH}
2H\delta H=\frac1{3M_P^2}[\mathcal{L}_{X}c_{s}^{-2}u_0\delta u+(V_{\phi}+\frac13V_{\phi}b)\delta\phi +\frac43T_0\delta s],
\end{equation}
can be obtained by varying the Friedmann equation. Then considering the definition of new slow roll parameters, we can work out the variations of $V_{\phi}$, $\Gamma$ and $r$ etc in terms of the fundamental perturbation variables:
\begin{equation}\label{deltagamma}
\frac{\delta \Gamma }{\Gamma _0}=\left( \frac 1{M_p^2}\frac{V_0}{V_\phi }\mathcal{L}_X
\beta +\frac c{3s_0T_0}V_\phi b\right) \delta \phi +\frac c{3s_0}\delta s,
\end{equation}
\begin{equation}\label{deltav}
\frac{\delta V_\phi }{V_\phi }=\left( \frac{V_{\phi\phi}}{V_{\phi}}+
\frac{V_\phi }{3s_0T_0}b^2\right) \delta \phi +\frac b{3s_0}\delta s,
\end{equation}
\begin{eqnarray}\label{deltar}
\frac{\delta r}{r_0} &=&\left[ \frac 1{M_p^2}\frac{V_0}{V_\phi }\mathcal{L}_X\beta+
\frac{bcV_\phi }{3s_0T_0}-\frac1{2V}(V_{\phi}+\frac13V_{\phi}b)\right] \delta
\phi   \nonumber \\
&&-\left(\frac1{2V}\mathcal{L}_X c_s^{-2}+\frac{\mathcal{L}_{XX}}{\mathcal{L}_X}\right)u_0\delta u+\left( \frac c{3s_0}-\frac{2T_0}{3V}\right) \delta s.
\end{eqnarray}
Above equations will be used repeatedly thereinafter.

Taking the variation of the Eqs. (\ref{EOM4}) and (\ref{entropy2}), we can obtain:
\begin{equation}  \left(
\begin{array}{c} \delta \dot \phi \\ \delta \dot u \\  \delta \dot s
\end{array} \right)  =E\cdot \left(\begin{array}{c} \delta \phi \\
 \delta u \\  \delta s \end{array} \right) -F.
\end{equation}
The $E$ is a $3\times3$ matrix which can be expressed as:
\begin{equation}
E=\left(\begin{array}{ccc} 0&1&0\\ A&\lambda_1 &B\\ C&D&\lambda_2
\end{array} \right).
\end{equation}
The matrix elements of $E$ can be calculated out:
\begin{eqnarray}\label{A}
A=3H_0^2c_s^2\left[\frac1{1+r}\epsilon-\eta+
\frac r{1+r}\beta -\frac{(1+r)^2}{r}b^2\right.\nonumber\\ \left.+(1+r)bc+\frac13\frac{1}{1+r}b\epsilon\right],~~~~~~~~~
\end{eqnarray}
\begin{equation}\label{B}
B=\frac{H_0T_0}{\mathcal{L}_X u_0}c_{s}^{2}\left[-c-\frac43\frac{\epsilon}{(1+r)^2}-\frac{1+r}{r}b\right],
\end{equation}
\begin{eqnarray}\label{C}
C=\frac{3H_0^2\mathcal{L}_Xu_0}{T_0}\left[\frac{r}{1+r}\epsilon-\frac{r}{1+r}\beta
\right.\nonumber\\ \left.+(1+r)(1-c)b+\frac13\frac{r}{1+r}b\epsilon\right],
\end{eqnarray}
\begin{equation}\label{D}
D=\frac{H_0\mathcal{L}_Xu_0}{T_0}\left[6r-\frac{rc_{s}^{-2}}{(1+r)^2}\epsilon\right],
\end{equation}
\begin{equation}\label{lambda1}
\lambda_1=-3H_0\left[1-(2+r)c_s^2-\frac23\mathcal{S}-\frac{1+c_s^2+rc_s^2}{3(1+r)^2}\epsilon\right],
\end{equation}
\begin{equation}\label{lambda2}
\lambda_2=H_0(c-4)-\frac43H_0\frac{r\epsilon}{(1+r)^2}.
\end{equation}
The column matrix $F$ acts as a small ``forcing term'', which can be expressed as:
\begin{equation}\label{F}
  F=\left(\begin{array}{c} 0\\ \dot u_0 \\ \dot s_0 \end{array}
\right).
\end{equation}
The parameter $\mathcal{S}$ in Eq. (\ref{lambda1}) is a characteristic quantity in noncanonical inflation expressed as $\mathcal{S}=\frac{\dot{c_s}}{Hc_s}$, describing the variation rate of the noncanonical effect.
The slow roll solutions can behave as an attractor for general warm inflationary dynamic system only when the eigenvalues of the matrix $E$ are negative or possibly positive but small quantities of order $\mathcal{O}(\frac{\epsilon}{1+r})$ (which means we can have slow growth), and the the ``forcing term'' $F$ is small enough, i.e. $|\frac{\dot
u_0}{ H_0u_0}|$ , $|\frac{\dot s_0}{H_0s_0}| \ll 1$.
Now we analyze the ``forcing term'' $F$ first. Taking the time derivative of the slow roll equations (\ref{EOM5}) and (\ref{entropy3}), and after some calculations, we get
\begin{equation}\label{dotu0}
\dot{u}_0=\frac{BC-A\lambda _2}{\lambda _1\lambda _2-BD}u_0,
\end{equation}
\begin{equation}\label{dots0}
\dot{s}_0=\frac{AD-C\lambda _1}{\lambda _1\lambda _2-BD}u_0.
\end{equation}
With the help of the matrix elements of $E$, we can finally calculate out
\begin{eqnarray}
\frac{\dot
u_0}{H_0u_0}&=&\frac1{\Delta}\left[\frac{4-c-rc}{1+r}\epsilon+\frac{4r}{1+r}\beta
+(c-4)\eta\right.\nonumber\\&+&\left.3(1+r)bc\right],
\end{eqnarray}
\begin{eqnarray}
\frac{\dot s_0}{H_0s_0}&=&\frac{1}{\Delta}\left[\frac{6-\mathcal{G}}{1+r}\epsilon+\frac{6+\mathcal{G}}{1+r}\beta
-6\eta\right.\nonumber\\ &+& \left.6(1+r)bc+\frac{1+r}{r}(c-1)\mathcal{G}b\right].
\end{eqnarray}
The quantity $\Delta$ in above two equations is $\Delta \simeq \mathcal{G}(c-4)+2rc$, where $\mathcal{G}=c_s^{-2}-(2+r)-\frac23\mathcal{S}c_s^{-2}$.

In slow roll inflationary epoch, the Hubble parameter should also be slowly varying, i.e. $\frac{\dot H_0}{H_0^2}\simeq-\frac{1}{\mathcal{L}_{X}(1+r)}\tilde{\epsilon}=-\frac{1}{1+r}\epsilon\ll1$.
Then the sufficient conditions characterized by new defined slow roll parameters to satisfy all above requirements can be obtained:
\begin{equation}\label{SR1}
    \epsilon\ll1+r,~~\beta\ll1+r,
    ~~\eta\ll r,~~b\ll\frac1{1+r};
  \end{equation}
when $c_s^2$ is not far less than unity;
and while $c^2_s\ll1$, we have
  \begin{equation}\label{SR2}
    \epsilon\ll c_s^{-2}+r,~~\beta\ll c_s^{-2}+r,
    ~~\eta\ll c_s^{-2}+r,~~b\ll\frac{c_s^{-2}+r}{r}.
  \end{equation}
Thus the validity of slow roll approximations can be measured by above slow roll conditions. The conditions are dimensionless and invariant under field transformations.
If we use the traditional slow roll parameters, above conditions can be expressed as:
\begin{equation}\label{SR3}
    \tilde{\epsilon}\ll\mathcal{L}_X(1+r),~~\tilde{\beta}\ll\mathcal{L}_X(1+r),
    ~~\tilde{\eta}\ll \mathcal{L}_Xr,~~b\ll\frac1{1+r};
  \end{equation}
when $c_s^2$ is not far less than unity;
and when $c^2_s\ll1$, we have
\begin{eqnarray}\label{SR4}
  \tilde{\epsilon}\ll \mathcal{L}_X(c_s^{-2}+r),~~~~\tilde{\beta}\ll \mathcal{L}_X(c_s^{-2}+r),\nonumber\\
    \tilde{\eta}\ll \mathcal{L}_X(c_s^{-2}+r),~~~~~~b\ll\frac{c_s^{-2}+r}{r}.
\end{eqnarray}
From above slow roll conditions, it is founded that the slow roll approximations we proposed in general noncanonical warm inflation is proper, and the validity of slow roll approximations can be easily guaranteed by the more relaxed slow roll conditions. Since the ``new'' slow roll parameters are invariant under field transformations, we can discuss the range of slow roll conditions in normalized field representation (the normalized field can always be get by making $\psi=\int\sqrt{\mathcal{L}_X}d\phi$). Then the parameter $\mathcal{L}_X$ is larger than one as discussed in Sec. \ref{sec:level2}, so the slow roll requirements on inflaton potential are really much more relaxed. The slow roll conditions in general noncanonical warm inflation are much broader than canonical warm inflation, let alone cold inflation, resulting from the strong noncanonical effect and thermal dissipation effect. Thus many kinds potentials that are excluded by cold inflation or even canonical warm inflation, can be consistent with noncanonical warm inflation. In ``coupling'' noncanonical warm inflation (i.e. the noncanonical field with a general Lagrangian density having coupling form of kinetic and potential terms), the slow roll approximations can be also given out, as in ``uncoupling'' noncanonical warm inflation (i.e. the noncanonical field with a somewhat easy Lagrangian density having separate form of kinetic and potential terms). The slow roll conditions suggest the noncanonical effect in general noncanonical warm inflation has more evident impact on the validity of slow roll inflation. The slow roll condition for parameter $b$ indicates thermal correction to the inflaton potential should be small, as in canonical warm inflation \cite{Ian2008,Campo2010}.

In the following, we analyze the matrix $E$ to give additional slow roll conditions. Using slow roll conditions we have obtained above, we reach the conclusion that
\begin{eqnarray}\label{charEq}
\det(\lambda I-E) &= &\left|
\begin{array}{ccc}
\lambda  & -1 & 0 \\
-A & \lambda-\lambda_1  & -B \\
-C & -D & \lambda-\lambda_2
\end{array}
\right|   \nonumber \\
&=&\lambda (\lambda-\lambda_1 )(\lambda-\lambda_2 )-BD\lambda-A(\lambda-\lambda_2)-BC   \nonumber
\\
&=&0
\end{eqnarray}
exists a very small eigenvalue $\lambda \simeq \frac{BC-A\lambda _2}{\lambda _1\lambda _2-BD-A}\ll\lambda_1,\lambda_2$. The other two eigenvalues are the solutions of the following equation:
\begin{equation}\label{lambda12}
 \lambda ^2-(\lambda _1+\lambda _2)\lambda +\lambda _1\lambda _2-BD=0.
\end{equation}
In order that the slow roll solutions can be an attractor for general warm inflationary dynamic system, the two eigenvalues should be both negative, which implies that $\lambda _1+\lambda _2<0$ and $\lambda
_1\lambda _2-BD>0$. Finally we get
 \begin{equation}
|c|<4.  \label{c}
\end{equation}

According to the slow roll conditions we have got above, the relation $\frac{\rho_{r}}{V}=\frac{r\epsilon}{2(1+r)^2}\ll 1$ is hold, which indicates radiation energy density is subdominated during the slow roll inflationary epoch. The conclusion is consistent with the slow roll requirement that the inflation is potential dominated.
Compared to cold inflation and canonical warm inflation, the slow roll approximations are easily to be guaranteed under the condition of noncanonical warm inflation.

\section{\label{sec:level4} A concrete example of general warm inflation}
Now we begin with a kinetic and potential coupling noncanonical warm inflationary model with the Lagrangian density:
\begin{equation}\label{DBI1}
\mathcal{L}(X',\varphi)=\Lambda^4\left[1-\sqrt{1-\Lambda^{-4}\varphi^2X'}\right]-V(\varphi).
\end{equation}
Obviously, it has a ``coupling'' form with $\mathcal{L}_{X'\varphi}\neq0$. In order to eliminate the coupling term, we define a new field $\phi=\frac{1}{2\sqrt{2}}\varphi^2$, then using the $\phi$ representation, the Lagrangian density becomes:
\begin{equation}\label{DBI2}
\mathcal{L}(X,\phi)=\Lambda^4\left[1-\sqrt{1-2\Lambda^{-4}X}\right]-V(\phi).
\end{equation}

Then the new Lagrangian density has the DBI form with a constant warp factor $f=\Lambda^{-4}$ as in \cite{Franche2010}, which is decoupled. For convenience, we shall consider the strong regime of warm inflation with $r\gg1$. The dissipation coefficient is assumed to have the form $\Gamma=C_{\phi}\frac{T^3}{\phi^2}$ as indicated in \cite{Mar2007}. The quantities associate with the strength of noncanonical effect is $\mathcal{L}_X=c_s^{-1}=\frac{1}{\sqrt{1-2fX}}>1$.

There are two kinds of DBI inflationary models: named the ultraviolet
(UV) model and the infrared (IR) model \cite{IRDBI,IRDBI1}. The UV DBI model is already at odds with observations, so we consider only the IR DBI model here. The IR DBI potential $V(\phi)$ in Eq. (\ref{DBI2}) can be expressed as $V(\phi)=V_0-\frac12\alpha H^2\phi^2$ \cite{IRDBI,IRDBI1}.
Then the slow roll conditions are given in the $\phi$ representation:
\begin{eqnarray}\label{DBISR}
\epsilon=\frac{\tilde{\epsilon}}{\mathcal{L}_X}=\frac{\alpha^2\phi^2}{18M_p^2\mathcal{L}_X}\ll r;\nonumber\\
\eta=\frac{\tilde{\eta}}{\mathcal{L}_X}=-\frac{\alpha}{3\mathcal{L}_X}\ll r;\nonumber \\ \beta=\frac{\tilde{\beta}}{\mathcal{L}_X}
=\frac{2\alpha}{3\mathcal{L}_X}\ll r.
\end{eqnarray}
Just meeting the conditions that the parameter $\alpha$ is not too large and the magnitude of inflaton is sub-Planckian, the slow roll conditions can be easily hold.
We have proposed a necessary criterion for a successful warm inflationary model that the Hubble slow roll (HSR) parameter $\epsilon_H$ must be an increasing function of time during the inflationary phase, thus $\epsilon_H$ can increase to 1 to end the inflation naturally and the Universe can turn smoothly into the radiation dominated phase \cite{TachyonZhang}. A crucial requirement for a warm inflationary model is having an increasing HSR parameter $\epsilon_H$. In this IR DBI warm inflationary model, the Hubble slow roll parameter $\epsilon_H=-\frac{\dot H_0}{H_0^2}\simeq\frac{\epsilon}{r}=
\frac{\alpha^2\phi^2}{18M_p^2\mathcal{L}_Xr}\propto\phi^4$. Along with the proceeding of inflation, the inflaton field is increasing, so we can have an increasing $\epsilon_H$, which is also consistent with the criterion. The conclusion can be reached that the IR DBI warm inflation model we assumed here is an practicable model.

\section{\label{sec:level5}Conclusions}

We review kinds of inflationary theories, and introduce the theory of general noncanonical warm inflation. General noncanonical warm inflation generalize the scope the inflation. We give the basic equations of this scenario, and review that an inconvenient ``coupling'' noncanonical model in $\varphi$ representation can turn to an easy-to-use ``uncoupling'' model in $\phi$ representation through field transformation $\phi=f(\varphi)$. We introduce the slow roll approximations and new slow roll parameters in the new scenario. The new slow roll parameters are all dimensionless as in canonical inflation, which have different form with the traditional definition of slow roll parameters. The main task of this paper is performing a systemic stability analysis to give out that under which conditions the slow roll solutions can really act as an attractor for the inflationary system. Then we get the slow roll conditions assuring the validity of slow roll approximations of the general warm inflationary picture which expressed by several slow roll parameters. The sound speed $c_s$ , measuring the strength of noncanonical effect, is invariant under field transformations. The slow roll conditions we got imply that slow roll inflation can still be tenable in general noncanonical warm inflation. The noncanonical effect resulting from kinetic and potential ``coupling'' will not destroy the definition of slow roll. What's more, the slow roll conditions are more relaxed compared to canonical warm inflation, let alone cold inflation, which is due to the enhanced Hubble damped term and thermal damped term in motion equation. The differences of conditions on $b$ and $c$ in canonical and noncanonical warm inflation are proved to be not huge, for they only describe the temperature dependence. Then we study the IR DBI warm inflationary model as a concrete example in general noncanonical warm inflation. We decouple the inflationary model by using an easy field transformation, calculate its slow roll conditions and a necessary warm inflationary criterion. And then we reach the conclusion that the choice of potential and dissipation coefficient etc. are suitable. This model can be a successful and self-consistent noncanonical warm inflationary model.

\acknowledgments This work was supported by the National Natural Science Foundation of China (Grant No. 11605100, No. 11704214, and No. 11975132).

\end{document}